\documentclass{bioinfo}
\copyrightyear{200x}
\pubyear{200x}

\usepackage{algorithm,amsmath}
\usepackage{algorithmic}

\begin{document}
\firstpage{1}

\title[Optimal flux states, reaction replaceability and response to knockouts in the hRBC]{Optimal flux states, reaction replaceability and response to knockouts in the human red blood cell}
\author[De Martino \textit{et~al}]{Andrea De Martino\,$^{1,2}$\footnote{to whom correspondence should be addressed}, Daniele Granata\,$^{2}$, Enzo Marinari\,$^2$, Carlotta Martelli\,$^3$ and Valery Van Kerrebroeck\,$^2$}
\address{$^{1}$CNR/INFM Centre for Statistical Mechanics and Complexity (SMC), Roma (Italy)\\
$^{2}$Dipartimento di Fisica, Sapienza Universit\`a di Roma, piazzale Aldo Moro 2, 00185 Roma (Italy)\\
$^3$Department of Molecular, Cellular and Developmental Biology, Yale University, New Haven, CT 06520 (USA)}

\history{~}%Received on XXXXX; revised on XXXXX; accepted on XXXXX}

\editor{~}%Associate Editor: XXXXXXX}

\maketitle

\begin{abstract}

\section{Motivation:}
Characterizing the capabilities, criticalities and response to perturbations of genome-scale metabolic networks is a basic problem with important applications. A key question concerns the identification of the potentially most harmful knockouts. The integration of combinatorial methods with sampling techniques to explore the space of viable flux states may provide crucial insights on this issue. 
\section{Results:}
We assess the replaceability  of every metabolic conversion in the human red blood cell by enumerating the alternative paths from substrate to product, obtaining a complete map of the potential damage of single enzymopathies. Sampling the space of optimal flux states in the healthy and in the mutated cell reveals both correlations and complementarity between topologic and dynamical aspects. 
\section{Availability:} 
Related material (stoichiometric matrices, C codes, optimal flux configurations and the detailed structure of alternative paths) is available from http://chimera.roma1.infn.it/SYSBIO

\section{Contact:} \href{andrea.demartino@roma1.infn.it}{andrea.demartino@roma1.infn.it}
\end{abstract}

\section{Introduction}

Understanding metabolic activity from the underlying genotype is one of the most addressed problems in computational biology. Functional modularity of metabolic networks suggests that topological aspects may provide a key to identify a class of `critical' structures or `essential' metabolic pathways \cite{maha,samal}. However the metabolic genotype only constitutes the frame on the top of which the dynamic phenotype is built. Criticality of a metabolic pathway will in general depend on both the network reconstruction from genomic information and the `model of metabolism' defined on it. In {\it E.coli}, phenotypical essentiality of metabolic genes has been associated with a reduced allowed variability of the corresponding fluxes, suggesting that dynamically stiff reactions may constitute an evolutionarily robust backbone of metabolism conserved over different species \cite{Mar}. 

Here we attempt a more thorough integration of topological and dynamical views that will provide further insights into the efficiency, robustness and responsiveness of a metabolic network. We will first associate the criticality of a reaction with its topological replaceability by enumerating the alternative paths from substrate to product along the network edges. Then we will validate and compare these results with the metabolic phenotype that results from the definition of a general constraint based model for metabolic flux prediction.

We carry out our analysis on the metabolic network of the human red blood cell (hRBC), one of the most studied organisms in systems biology, from the earliest mathematical models of single biochemical pathways \cite{rapo,holz} to the currently available genome-scale reconstructions \cite{sysbio}. The reason for this choice lies essentially in its limited size. On one hand, it allows to compute reaction replaceabilities by a suitable modification of Johnson's exact algorithm for counting loops in a directed graph \cite{Joh}. On the other, it allows for the efficient application of various sampling methods to the space of viable flux states \cite{Wib,Mar}. This is vital to address many important properties of erythrocytes. Indeed for some organisms under certain conditions it is reasonable to assume that the metabolic activity is aimed at maximizing a subset of the metabolic reactions (or a function of them) associated with a certain biological function. In such cases the relevant flux configuration can be computed by standard optimization algorithms. For example, {\it E.coli}'s metabolism has been shown to maximize biomass production under evolutionary pressure \cite{edw}, but after a genetic knockout it responds with a minimum rearrangement of fluxes \cite{seg}. While the production of the cofactors ensuring the maintenance of osmotic balance and the release of oxygen may be argued to be their metabolic goal, erythrocytes do not generically allow for such a simplification. Information-rich directions in flux space must be retrieved by coupling the underlying constraints on fluxes with other types of analyses. Much understanding has indeed been obtained from the uniform sampling of feasible states \cite{Wib,Pri,Bra} and by functional studies, like the computation of extreme pathways \cite{wiba}, of metabolic regulatory structures \cite{pricea,barrett} and of metabolic pools \cite{kau2}. 

These aspects combined make hRBCs a key benchmark for both theories of metabolism and computational tools.

\section{Approach}

\begin{figure}[t]
\centering
\includegraphics[width=5cm]{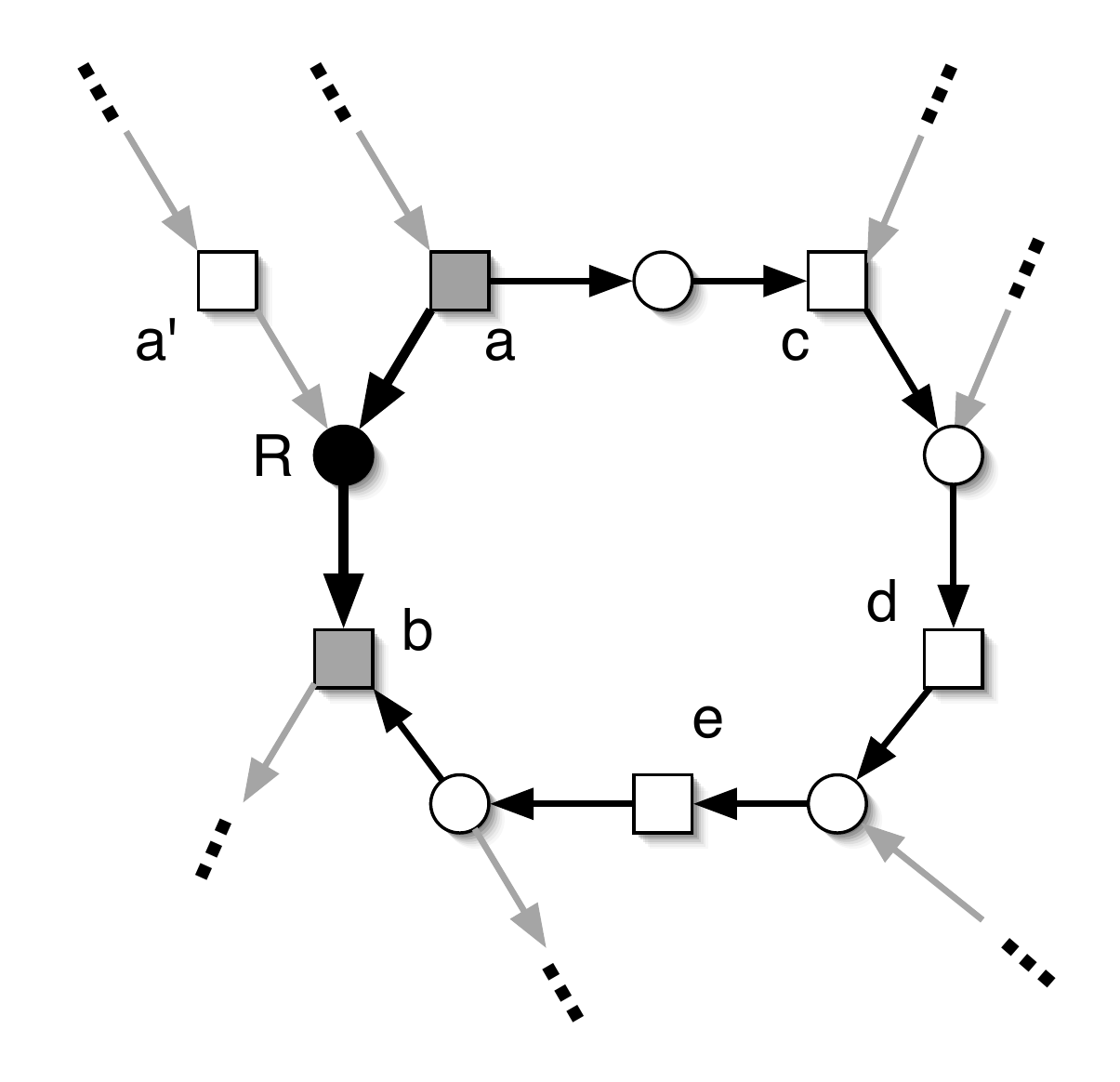}
\caption{\label{loop}Bipartite graph representation of a reaction network, with circles (resp. squares) denoting reactions (resp. metabolites). Here, reaction $R$ uses metabolites $a$ and $a'$ as substrates to produce metabolite $b$. If $R$ is knocked out, the conversion of $a$ to $b$ is still permitted by the alternative pathway $a\to c\to d\to e\to b$. When $R$ is fictitiously reversed, this chain forms a directed loop of length $5$ reactions, formed by $R$ and by a path passing through $\ell=4$ other reactions.}
\end{figure}

Given a reaction network, we want to compute, for any pair of metabolites $a$ and $b$ that are respectively substrate and product in a reaction $i$ (this situation will be indicated by $a\!\!\overset{i}{\to} \!\!b$), the number $\mathcal{N}_{a\to b}^{(i)}(\ell)$ of alternative pathways, excluding reaction $i$, of length $\ell$ allowing for the conversion $a\!\!\to\!\! b$, see Fig.\ref{loop}. 
The idea is that a metabolite conversion $a\!\!\overset{i}\to\!\! b$ for which $\mathcal{N}_{a\to b}^{(i)}(\ell)$ (or, more properly,$\sum_\ell\mathcal{N}_{a\to b}^{(i)}(\ell)$) is large will be more easily substituted in case of a knockout than one for which the above quantity is small. Finding paths connecting two points of a directed network is a long-studied problem in computer science. The focus is usually on locating the shortest paths or the fastest way to find any path.  Characterizing all the distinct paths between two vertices is however a less confronted issue. In our case it is crucial to avoid overcounting, e.g. due to self-intersecting paths. Therefore we shall resort to an exhaustive algorithm. We will identify the substitutive paths using the following trick: for each pair $(a,b)$ of metabolites such that $a\!\!\overset{i}{\to}\!\!b$, revert $i$ fictitiously. This results in a new graph where an auxiliary edge $b\!\!\overset{i}{\dashrightarrow}\!\!a$ replaced the edge $a\!\!\overset{i}{\to}\!\!b$, see again Fig.\ref{loop}. Counting the number of alternative reaction chains producing $b$ from $a$ then comes down to computing the number of directed cycles, i.e., non self-intersecting directed closed paths along the  edges of the new graph, passing through the fictitious edge $b\overset{i}{\dashrightarrow}a$. Thanks to the limited size of the hRBC network it is possible to solve this enumeration problem exactly via Johnson's algorithm \cite{Joh}, briefly described in the following section. $\mathcal{N}_{a\to b}^{(i)}(\ell)$ can now be trivially inferred. For simplicity, $\ell$ will denote here the number of reactions in the alternative pathway ($\ell=4$ in Fig.\ref{loop}).

The space of viable fluxes will be defined through a constraint-based approach which relies on more general assumptions than flux-balance analysis (FBA, \cite{Kau}). FBA is the standard method to model steady-state reaction networks where mass balance constraints are imposed to every metabolite. 
For a reaction network with $N$ reactions and $M$ metabolites, let us denote by $\boldsymbol{A}$ and $\boldsymbol{B}$, respectively, the $M\times N$ matrices of output and input stoichiometric coefficients. The stoichiometric matrix is given by $\boldsymbol{S}=\boldsymbol{A}-\boldsymbol{B}$. Letting $\boldsymbol{\nu}=(\nu_i)_{i=1}^N$ denote a vector of fluxes (with properly chosen bounds  $ \nu_i^{min}\leq \nu_i\leq \nu_i^{max}$), the concentrations $\boldsymbol{c}=(c^a)_{a=1}^M$ of metabolites vary in time according to $\dot{\boldsymbol{c}}=\boldsymbol{S\nu}-\boldsymbol{u}$, where $\boldsymbol{u}=(u^a)_{a=1}^M$ stands for the net cellular uptake of metabolite $a$ ($u^a>0$ if $a$ is a global output of metabolism, $u^a<0$ if $a$ is consumed by the organism, $u^a=0$ if $a$ is mass-balanced). Assuming a steady state, the concentrations are constant in time (i.e. $\dot{\boldsymbol{c}}=0$) and vectors $\boldsymbol{\nu}$ satisfying $\boldsymbol{S\nu}=\boldsymbol{u}$, or 
\begin{equation}\label{fba}
\boldsymbol{(A-B)\nu}=\boldsymbol{u}, 
\end{equation}
represent flux configurations ensuring that each metabolite meets its production or consumption constraints at fixed concentrations. As $N$ is typically larger than $M$, the system is underdetermined and feasible flux states form a convex set of dimension $N-{\rm rank}(\boldsymbol{S})$ embedded in the $N$-dimensional space of fluxes. In absence of a selection criterion that allows to pick one solution out of this set (as e.g. a maximum biomass principle), a uniform sampling of the solution space should be carried out. This can be achieved effectively, albeit at a considerable computational cost, by Monte Carlo methods \cite{Wib, Pri} or by message-passing procedures \cite{Bra}.

Here we will consider a different but related approach based on Von Neumann's (VN) model of reaction networks \cite{Mar}. In the VN framework we fix the environment through a small set of intakes on nutrients and define a self-consistent flux problem where the network chooses, given a target growth rate, how much of the nutrients to use and which metabolites are globally produced. Mass balance then emerges as a property of the solutions for some metabolites. The equations describing the VN model have been studied by statistical mechanics methods in \cite{DeMM,DeM}. For an intuitive derivation, note that the quantities $\boldsymbol{A\nu}$ and $\boldsymbol{B\nu}$ represent, respectively, the total output and the total input of each metabolite for a given flux vector $\boldsymbol{\nu}$. Then a flux vector such that $\boldsymbol{A\nu}\geq\rho\boldsymbol{B\nu}$, with some constant $\rho>0$, describes a network state where metabolites are being produced at a rate at least equal to $\rho$, since for each of them the total output is at least $\rho$ times the total input.  It is simple to see that as $\rho$ increases the volume of such flux vectors shrinks continuously (for $\rho=0$ every flux vector is a solution). In particular, there exists a value $\rho^\star$ of $\rho$, representing the maximum metabolic production rate compatible with the stoichiometric constraints, above which no suitable flux vectors exist. The presence of conserved metabolic pools \cite{Fam} implies $\rho^\star=1$ \cite{DeM2}, so that in metabolic networks optimal fluxes correspond to the solutions of
\begin{equation}\label{vn}
\boldsymbol{(A-B)\nu}\geq\boldsymbol{0}
\end{equation}
The solutions of (\ref{vn}) do not coincide with those of (\ref{fba}) even for $\boldsymbol{u}=\boldsymbol{0}$. Interestingly, a finite volume of (optimal) flux states turns out to satisfy the above constraints \cite{Mar}. This trait is at odds with both the behavior of the solutions of (\ref{vn}) for a random reaction network (where a single solution survives at $\rho^\star$ \cite{DeM}) and with the optimization that is usually coupled to FBA (where typically a single flux state maximizes the objective function), and points to the robustness of metabolic phenotypes. For {\it E.coli}, in particular, the solutions of (\ref{vn}) have been shown to reproduce both the large-scale organization of fluxes and the individual measured rates. In addition, fluxes with the smallest solution-to-solution fluctuations, representing the most susceptible parts of the network, turn out to be strongly correlated with {\it E.coli}'s phenomenologically essential genes \cite{Mar}.

\vspace{-0.5cm}

\begin{methods}
\section{Methods}

\subsection{Reconstructed network}

We consider the hRBC metabolic network studied in \cite{Pri}, a map of which is shown in Fig. \ref{mappa}; Table \ref{tab} lists reactions and the corresponding abbreviations. 
\begin{figure}[t]
\centering
\includegraphics[width=7.5cm]{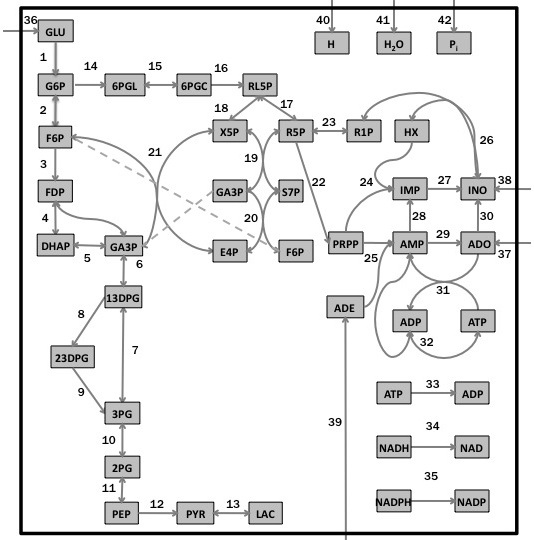}
\caption{\label{mappa}Scheme of the hRBC metabolic network used in our analisys. Squares correspond to metabolites, numbers to reactions (see Table \ref{tab}).}
\label{fig:map}
\end{figure} 
\begin{table}
\begin{center}
\begin{tabular}{lll}
\hline 
N. & Abbr. &  Chemical reaction \\
\hline
1 &HK &  GLU + ATP $\rightarrow$ G6P + ADP + H \\
2 &PGI &  G6P $\leftrightarrow$ F6P\\
3 &PFK & F6P + ATP $\rightarrow$ FDP + ADP + H\\
4 &ALD& FDP $\leftrightarrow$ GA3P + DHAP\\
5 &TPI &  DHAP $\leftrightarrow$ GA3P\\
6 &GAPDH   & GA3P+NAD+P$_i\leftrightarrow$13DPG+NADH+H\\
7 &PGK & 13DPG + ADP $\leftrightarrow$ 3PG + ATP\\
8 &DPGM & 13DPG $\rightarrow$ 23DPG + H\\
9 &DPGase & 23DPG + H2O $\rightarrow$ 3PG + P$_i$\\
10 &PGM & 3PG $\leftrightarrow$ 2PG\\
11 &EN & 2PG $\leftrightarrow$ PEP + H2O\\
12 &PK & PEP + ADP + H $\rightarrow$ PYR + ATP\\
13 &LDH  & PYR + NADH + H $\leftrightarrow$ LAC + NAD\\
14 &G6PDH  & G6P + NADP $\rightarrow$ 6PGL + NADPH + H\\
15 &PGL & 6PGL + H$_2$O $\leftrightarrow$ 6PGC + H\\
16 &PDGH  & 6PGC + NADP $\rightarrow$ RL5P + NADPH + CO$_2$\\
17 &RPI &  RL5P $\leftrightarrow$ R5P\\
 18 &XPI & RL5P $\leftrightarrow$ X5P\\
19 &TKI & X5P + R5P $\leftrightarrow$ S7P + GA3P\\
20 &TA & GA3P + S7P $\leftrightarrow$ E4P + F6P\\
21 &TKII & X5P + E4P $\leftrightarrow$ F6P + GA3P\\
22 &PRPPsyn &  R5P + ATP $\rightarrow$ PRPP + AMP\\
23 &PRM  & R1P $\leftrightarrow$ R5P\\
24 &HGPRT & PRPP + HX + H$_2$O $\rightarrow$ IMP + 2 P$_i$\\
25 &AdPRT  & PRPP + ADE + H$_2$O $\rightarrow$ AMP + 2 P$_i$\\
26 &PNPase  & INO + Pi $\leftrightarrow$ HX + R1P\\
27 &IMPase  & IMP + H$_2$O $\rightarrow$ INO + P$_i$ + H\\
28 &AMPDA & AMP + H$_2$O $\rightarrow$ IMP + NH$_3$\\
29 &AMPase  & AMP + H$_2$O $\rightarrow$ ADO + P$_i$ + H\\
30 &ADA &ADO + H$_2$O $\rightarrow$ INO + NH$_3$\\
31 &AK &ADO + ATP $\rightarrow$ ADP + AMP\\
32 &ApK & 2 ADP $\leftrightarrow$ ATP + AMP\\
33 &ATPase &    ATP $\rightarrow$  ADP + P$_i$\\
34 &NADHase   & NADH $\rightarrow$  NAD + H\\
35 &NADPHase & NADPH $\rightarrow$  NADP + H\\
%36 & GLUin &\\
%37 & ADEin &\\
%38 & INOin & \\
%39 & ADOin & \\
%40 & Hin & \\
%41 & H$_2$Oin & \\
%42 & Pin & \\
\hline
\end{tabular}
\end{center}
\caption{\label{tab}List of reactions considered in this work, including the corresponding number in the map of Fig.\ref{mappa}, the abbreviation and the process. The $7$ uptake fluxes, numbered 36 to 42, are as shown in Fig.\ref{mappa}.}
\end{table}
The network comprises three main pathways, namely glycolysis (reactions 1--13), the pentose phosphate (PP) pathway (14--21) and the adenosine metabolism, with a total of $M=39$ metabolites linked by $N=59$ reactions: 49 internal reactions (34 of which come from the splitting of 17 reversible processes), 3 auxiliary fluxes to maintain the osmotic equilibrium and the redox  state of the cell  (ATPase, NADHase, NADPHase) and 7 uptake reactions to guarantee the intake of the necessary nutrients (GLU, ADE, ADO, INO) and of the cytosol elements (H$_2$O, H, P$_i$).

\subsection{Structural properties}

Structural vulnerabilities are identified by analyzing the loop structure of a modified metabolic reaction network, created from the original one by inverting --in turn-- the direction of the single reaction for which we want to compute the replaceability, as explained in Fig.\ref{loop}. The fastest known exact algorithm (for the worst case scenario) of this cycle enumeration problem for a directed graph was introduced by Johnson \cite{Joh}. We shall now shortly describe its key ideas, referring to \cite{Joh} for a pseudocode. Given a directed graph with $n$ vertices and $e$ edges, the algorithm is designed to build non-selfintersecting paths from a root vertex $r$ to itself, loading them onto stacks. The main ingredients allowing for an optimal exploration of the graph are (a) a smart choice of the root vertex and (b) an efficient method to avoid duplicating cycles and repeating searches on the same portions of the graph. To achieve this, vertices are initially ordered in some lexicographic sequence, and the algorithm only selects as roots those nodes that are the ``least'' vertex (in the initial ordering) of at least one cycle. The algorithm described in \cite{Tarjan72} garantuees to find such vertices in $O(n+e)$ operations. Moreover, to avoid self-intersections, each time a node is loaded onto a stack it is also given a ``blocked'' status. It was proven by Johnson that if a vertex $v$ stays blocked as long as every path from $v$ to the root vertex $r$ intersects the current path at a vertex other than $r$, the algorithm outputs all cycles exactly once. By sufficiently delaying the unblocking of each of these vertices and by keeping track of the portions of the graph that have been searched holding the current stack, the maximum time that can elapse between two consecutive cycle outputs can be reduced to $O(n+e)$. The same holds for the time window before the first cycle is delivered and for the one after the output of the last cycle. Hence, the total time needed to list the, say, $c$ cycles of the graph is $O ((n+e)(c+1))$. In our case, each fictitious reaction reversal generates a new graph, so that computing the complete substitutability map for the hRBC requires a time of the order $O (N(n+e)(c+1))$. For practical reasons, we perform this analysis on the bipartite metabolic network (as in Fig.\ref{loop}) rather than the reduced network of Fig.\ref{mappa}. This implies that in our case $n=N+M$. 

One can in principle consider different measures of replaceability of a metabolic conversion $a\!\!\overset{i}{\to}\!\! b$. The quantity $\mathcal{R}_{a\to b}^{(i)}=\sum_\ell\mathcal{N}_{a\to b}^{(i)}(\ell)$, which counts the total number of paths alternative to $i$ from $a$ to $b$ of any length, is perhaps the most obvious option, while taking into account the fact that longer detours are less convenient than shorter ones from an energetic viewpoint one is lead to consider e.g. exponentially-weighted functions like $\mathcal{W}_{a\to b}^{(i)}=\sum_\ell \exp(-\ell)\mathcal{N}_{a\to b}^{(i)}(\ell)$. $\mathcal{R}$-based and $\mathcal{W}$-based rankings of metabolic conversion are rather different. The full rankings are available from http://chimera.roma1.infn.it/SYSBIO. Here we limit ourselves to identifying three key reaction groups which are independent of the replaceability measure: (a) the group of reactions such that each substrate-product pair involved in them can be substituted (this is putatively the part of the network that is most robust to single knockouts); (b) the group of reactions that cannot be substituted, corresponding to the most harmful enzymopathies; (c) the group of reversible reactions that are only replaceable in one direction, corresponding to the situation in which a conversion $a\leftrightarrow b$ can only be  substituted in one direction in case of a knockout. Note that, for topological reasons, intakes are not replaceable.

\subsection{Optimal flux configurations}

Optimal flux vectors, i.e. solutions of (\ref{vn}), are computed by the algorithm introduced in \cite{DeM} based on \cite{Kra}. The idea is to modify fluxes iteratively until all inqualities in (\ref{vn}) are satisfied. Specifically, for a fixed $0\leq \rho<\rho^\star$  (with $\rho^\star=1$ in our case) define $\boldsymbol{\Xi}(\rho)=\boldsymbol{A}-\rho \boldsymbol{B}$ and let $\boldsymbol{\xi}^a(\rho)$ denote the rows of $\boldsymbol{\Xi}(\rho)$, for $a\in\{1,\ldots,M\}$. Let also, for each iteration step $t$, $\boldsymbol{\nu}(t)$ be the flux vector at step $t$ and
\begin{equation}
m(t)=\text{arg~}\min_a ~\boldsymbol{\xi}^a(\rho)\cdot\boldsymbol{\nu}(t)
\end{equation}
At each $t$, the algorithm runs as follows. If $\boldsymbol{\xi}^{m(t)}(\rho)\cdot\boldsymbol{\nu}(t)<0$, update fluxes according to
\begin{equation}
\nu_i(t+1)=\max\{0,\nu_i(t)+\xi_i^{m(t)}(\rho)\}
\end{equation}
and iterate in $t$. Else, if $\boldsymbol{\xi}^{m(t)}(\rho)\cdot\boldsymbol{\nu}(t)\geq 0$ stop, i.e. $\boldsymbol{\nu}(t)$ is a solution.

Convergence to a solution is guaranteed for all $0\leq \rho<\rho^\star$ \cite{DeM}, so that $\rho^\star$ can be approximated with the desired resolution by iterating the above process for increasing values of $\rho$. To ensure that solutions are well defined one can either resort to setting fixed upper bounds on $\nu_i$'s or, as we do, impose a linear constraint of the form $\sum_i\nu_i(t)=N$ on the solutions (this is equivalent to singling out one flux as the reference unit for the other fluxes). It is convenient to initialize the algorithm with a random vector $\boldsymbol{\nu}(0)$. Different initial points generate trajectories to different solutions at $\rho^\star$ and the sampling of the solution space thus obtained turns out to be uniform \cite{Mar}.

As a means to characterize the shape of the solution space we employ the average overlap between different optimal flux vectors, defined as follows. Let $\boldsymbol{\nu}_\alpha$ and $\boldsymbol{\nu}_\beta$ denote two distinct solutions of (\ref{vn}) and, for each flux $i$, let 
\begin{equation}\label{ovl}
q_{\alpha\beta}(i)=\frac{2 \nu_{i\alpha}\nu_{i\beta}}{\nu_{i\alpha}^2+\nu_{i\beta}^2}
\end{equation}
This quantity, called the `overlap' between $\alpha$ and $\beta$, equals $1$ if flux $i$ takes on the same value in solutions $\alpha$ and $\beta$ and decreases as the values differ more and more. Averaging $q_{\alpha\beta}(i)$ over different pairs of solutions provides a measure of the allowed variability of flux $i$ (smaller variability corresponds to larger average overlap), complementary to the standard deviation of the resulting flux distribution. The complex shape of the solution space can then be roughly understood by distinguishing narrower directions with larger overlap or less variable fluxes from broader ones. It is reasonable to think that a cell will be more sensitive to perturbations (e.g. knockouts) of fluxes with larger overlap. Hence analyzing the susceptibility of the solution space to perturbations along the directions identified by different fluxes allows to extract a list of the potentially more deleterious perturbations, in analogy with previous work on {\it E.coli} \cite{Mar}.

\subsection{Response to enzymopathies}

In order to test the hRBC network against enzymopathies, we can focus on two types of perturbations. One can first employ a structural criterion: the knockout of a metabolic conversion $a\to b$ that is less ``substituted'' is more likely to be deleterious for the cell than the knockout of a highly replaceable conversion. The second criterion is based on fluxes: fluxes with smaller allowed variability (i.e. larger overlap) in the healthy cell are more likely to be critical links of the network than fluxes whose value can be changed over a larger range without losing optimality. As is to be expected, the knockouts rankings produced in these ways have a large degree of similarity, as reactions in the group (b) discussed above coincide with the physiologically most critical part of hRBC's metabolism. The simplest way to simulate an enzymopathy on flux $i$ is to constrain its value below a certain upper bound $\overline{\nu}_i$. Deficiencies can be partial, i.e., of a smaller degree, the closer $\overline{\nu}_i$ is to the upper limit of the allowed range in the healthy cell, or total if $\overline{\nu}_i=0$. Such constraints cause in principle a modification of the solution space along the direction $i$ which in turn cascades on the entire volume, modifying the optimal states of the metabolic network. 

\end{methods}

\section{Results}

\subsection{Structural properties}

\begin{figure}[!]
\centering
\includegraphics[width=7.5cm]{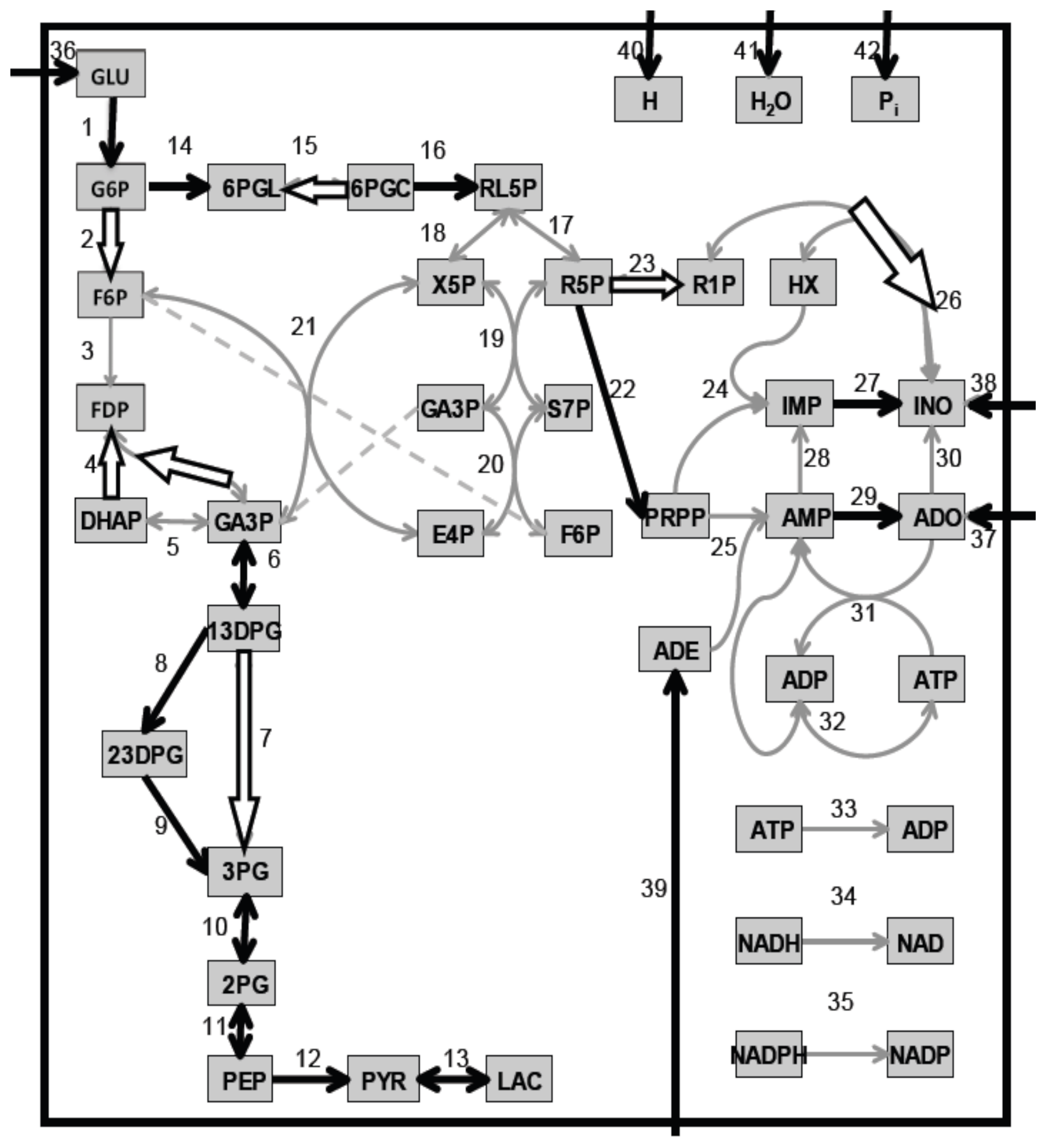}
\caption{\label{lupi}hRBC's reaction replaceability map. Black arrows denote unreplaceable reactions (group (b) above); thick white-filled arrows denote reversible reactions that can be replaced only in the direction indicated by the arrow (group (c)); all other reactions are fully replaceable (group (a)). }
\end{figure}

The substitutability map derived from the loop analysis is displayed in Fig. \ref{lupi}. (For the sake of simplicity we exclude the highly replaceable currency exchange fluxes from this discussion.)
The most replaceable core of the network lies in the PP pathway (reactions 17--21), which constitutes the main source of NADPH, a key metabolite in erythrocytes that limits the accumulation of peroxides protecting the cell from hemolysis. The high reliability coming with replaceability partly explains the reason why this group of reactions plays a central role not just as an auxiliary pathway for glycolysis, see the following analysis of fluxes. Unreplaceable reactions are instead lined up along glycolysis (numbers 1,6,8--13), in the bridge between glycolysis and the PP-pathway (14 and 16) or in auxiliary modules (22, 27, 29; the ADE$\to$AMP conversion in 25 is also not replaceable being directly linked to the ADE uptake). The physiologically most deleterious knockouts (HK, PK and G6PDH) all belong to this group. For instance, deficiency in the level of G6PDH is the basis of different types of hemolytic anemias, including favism, and is also linked to malaria resistance \cite{fm}. Finally, there is a group of reversible reactions that can be replaced only in one direction (reaction 2, 4, 7, 15, 23, 26). The last three of these could still be replaced in case of a knockout if a proper medium is selected. For instance, if reaction 15 is knocked out, it could be substituted by an alternative chain of reactions provided 6PGC is externally supplied. This is instead not possible for reaction 4 and possibly 23 (depending on the directionality of reaction 26), as a knockout in these cases would necessarily result in a net production of FDP and R1P.

\subsection{Optimal flux configurations}

\begin{figure*}%figure2
\centerline{\includegraphics[width=12cm]{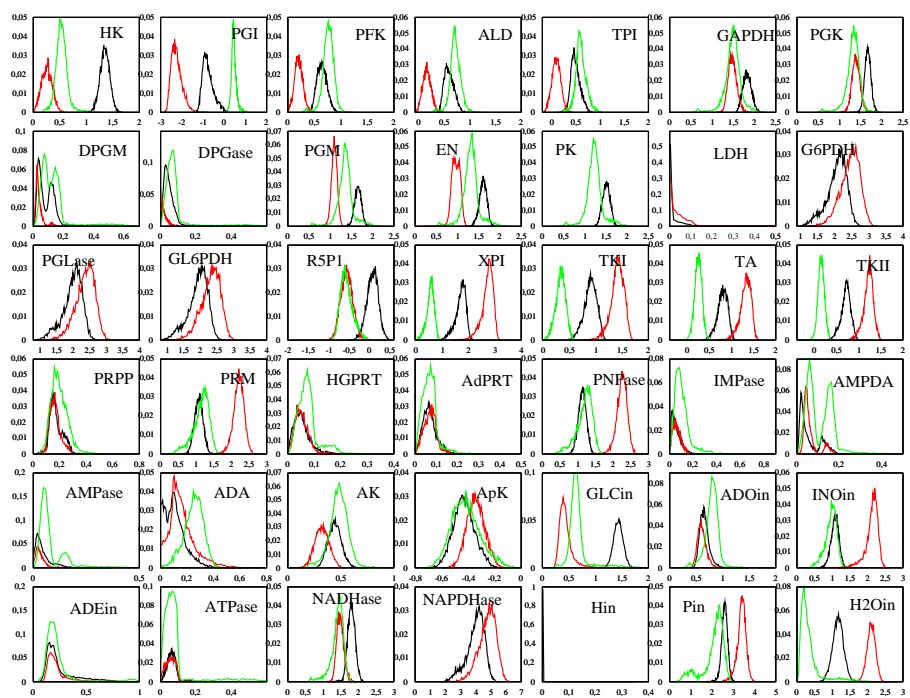}}
\caption{\label{dis}Distributions of differential fluxes (measured for each reaction by the difference between the direct and the reverse flux, when possible) for the healthy cell (black lines) and for the hRBC with knockout PK (red) e G6PDH (green).}
\end{figure*}

\begin{figure*}[t]
\centering
\includegraphics[width=7.5cm]{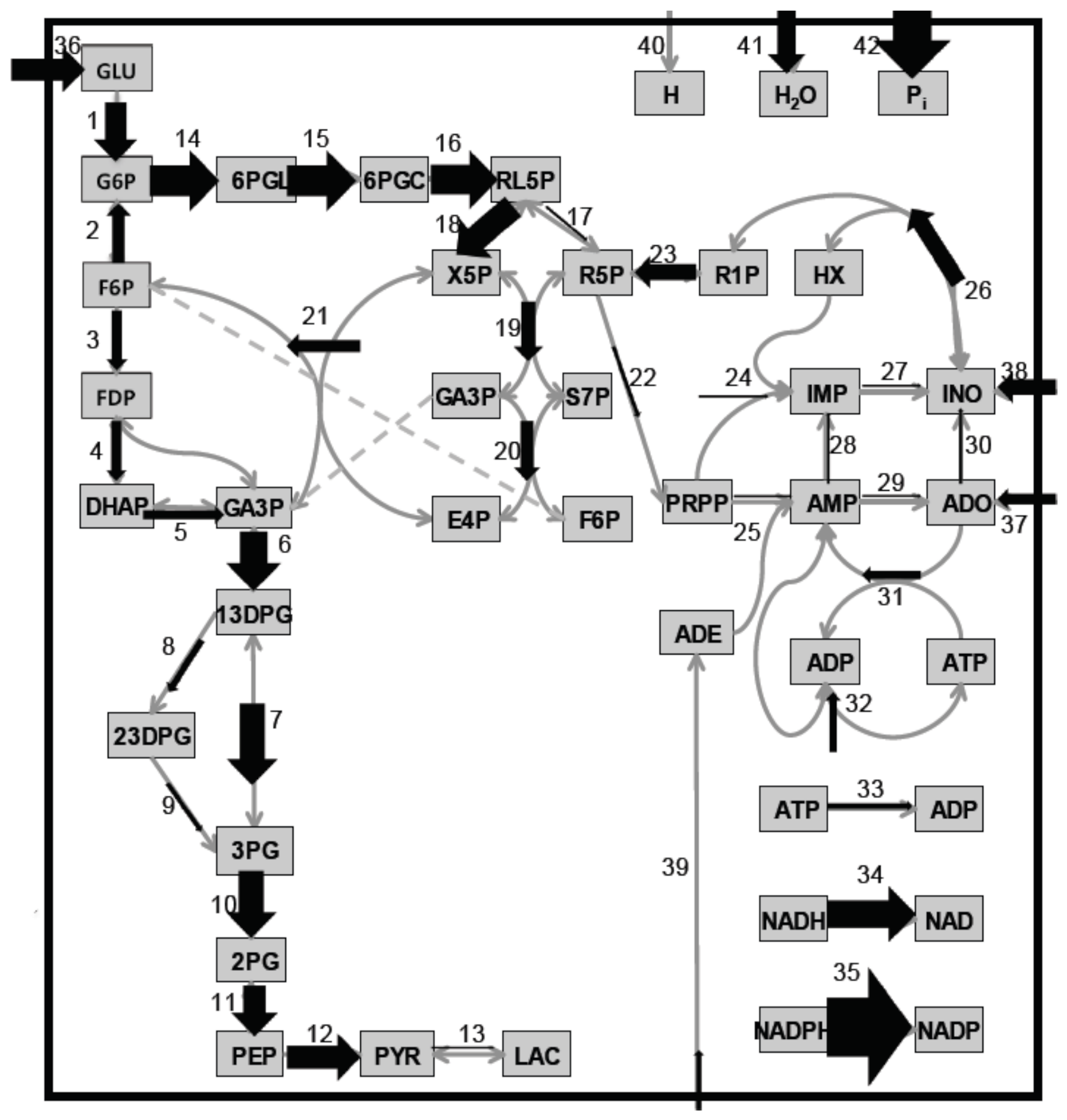}
\includegraphics[width=7cm]{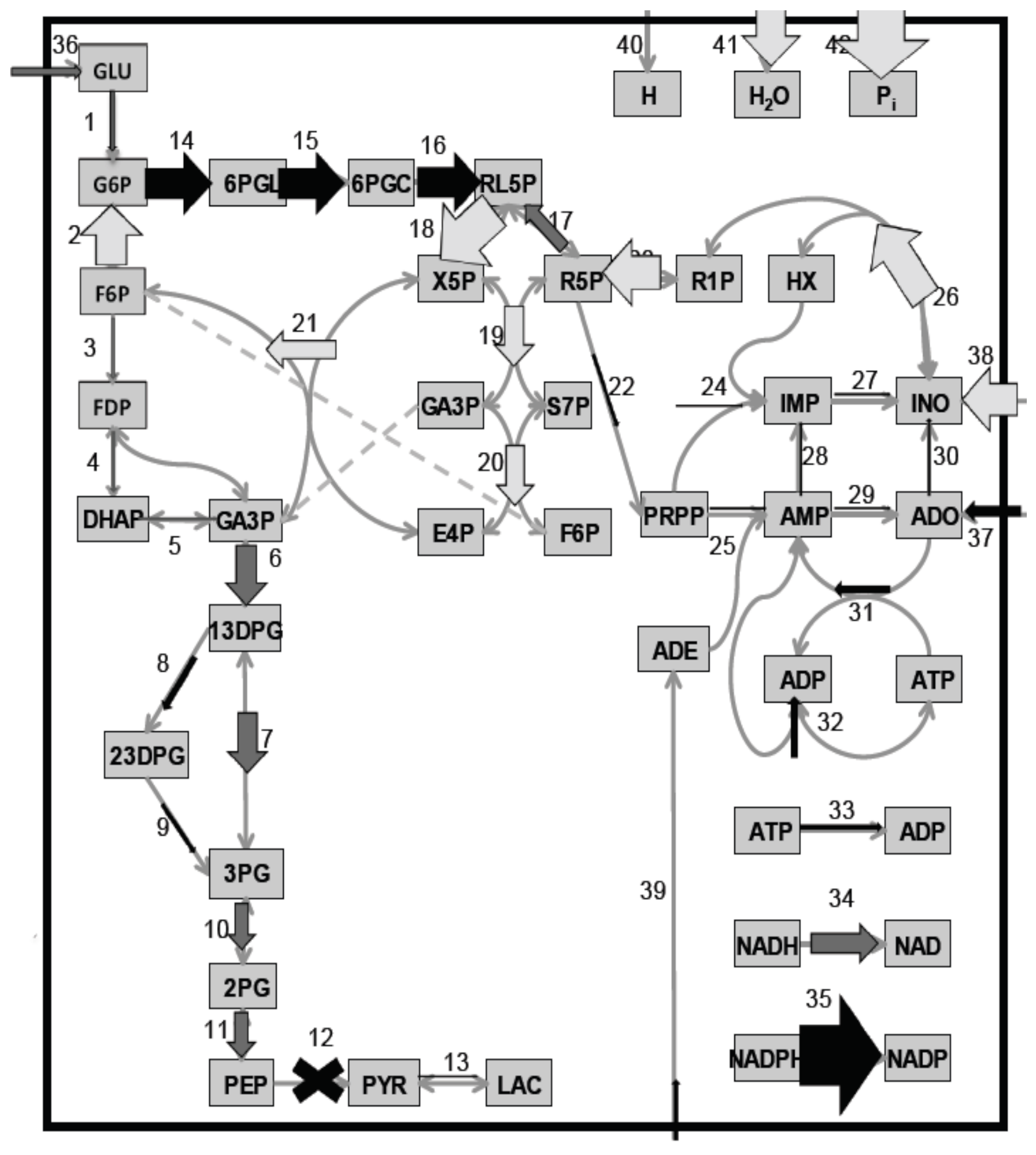}
\includegraphics[width=7cm]{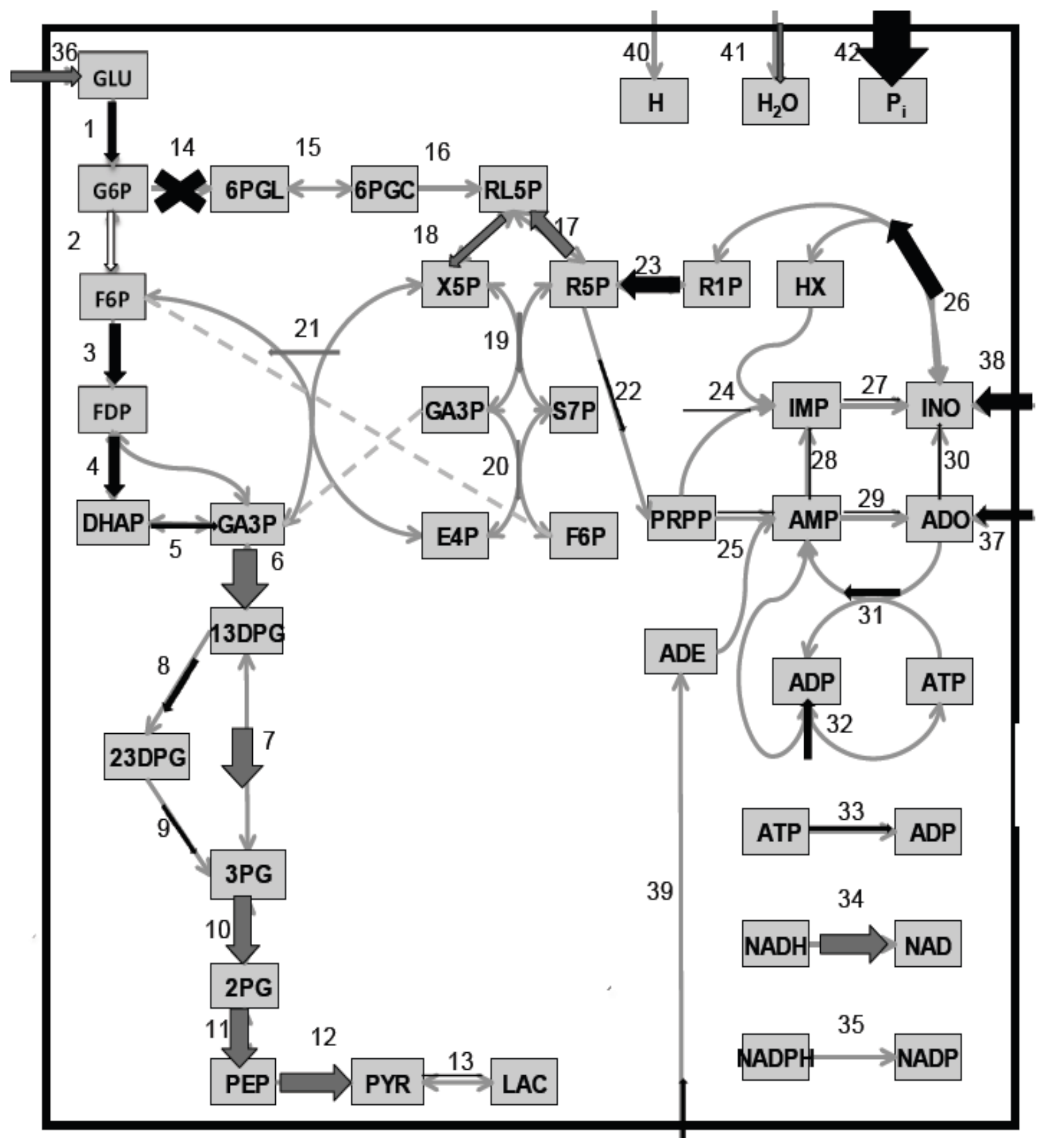}
\includegraphics[width=7cm]{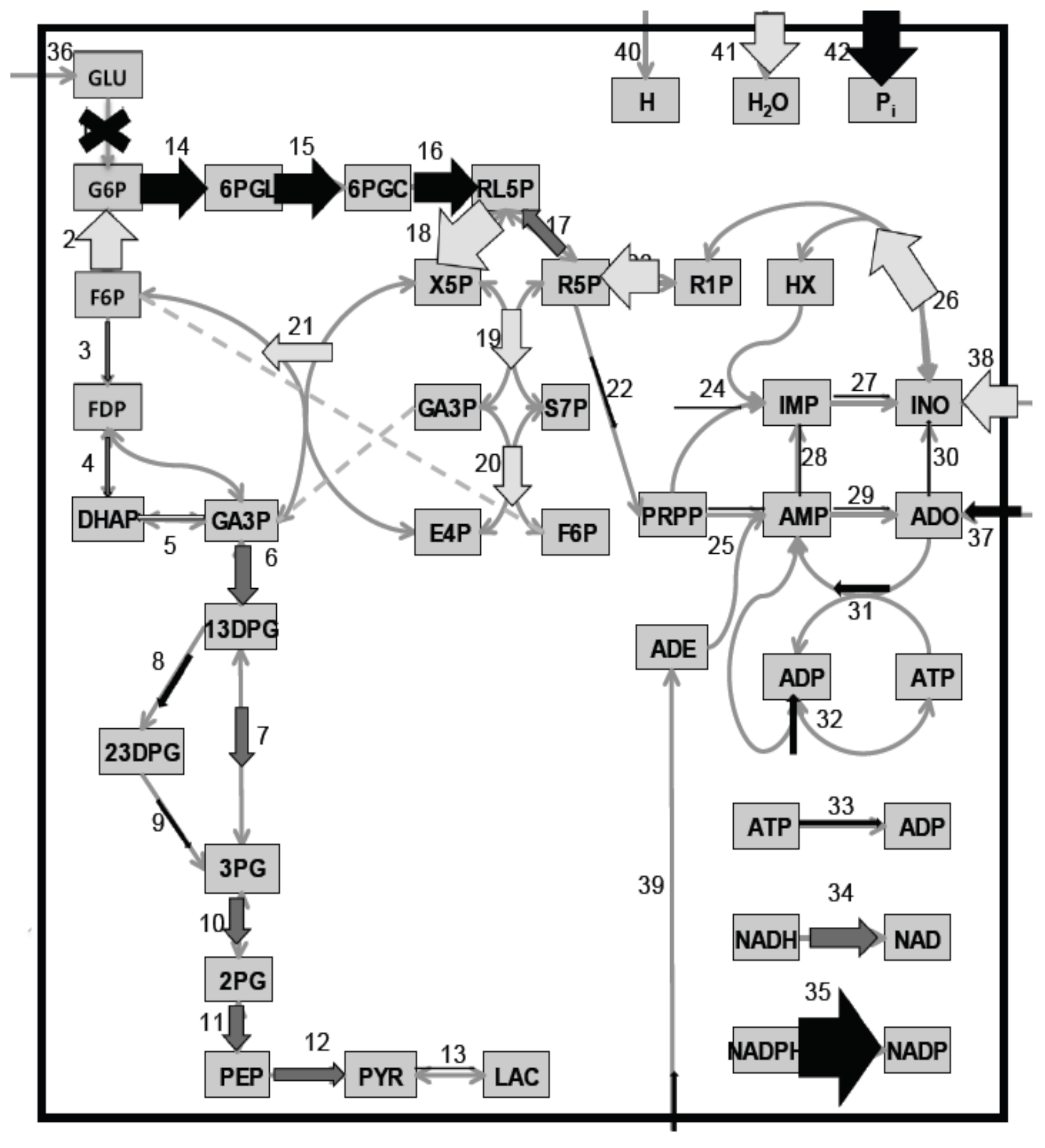}
\caption{\label{map}Maps of fluxes in the healthy hRBC (top left) and flux rerouting in three full enzymopathies: PK (top right), G6PDH (bottom left) and HK (bottom right). The thick arrows are dark grey if the reaction rates diminish with respect to the healthy cell, light grey if they are higher, black if they remains the same. The widths of the arrows are proportional to the mean value of the corresponding distributions.}
\end{figure*}

\begin{figure}%figure2
\centerline{\includegraphics[width=7.5cm]{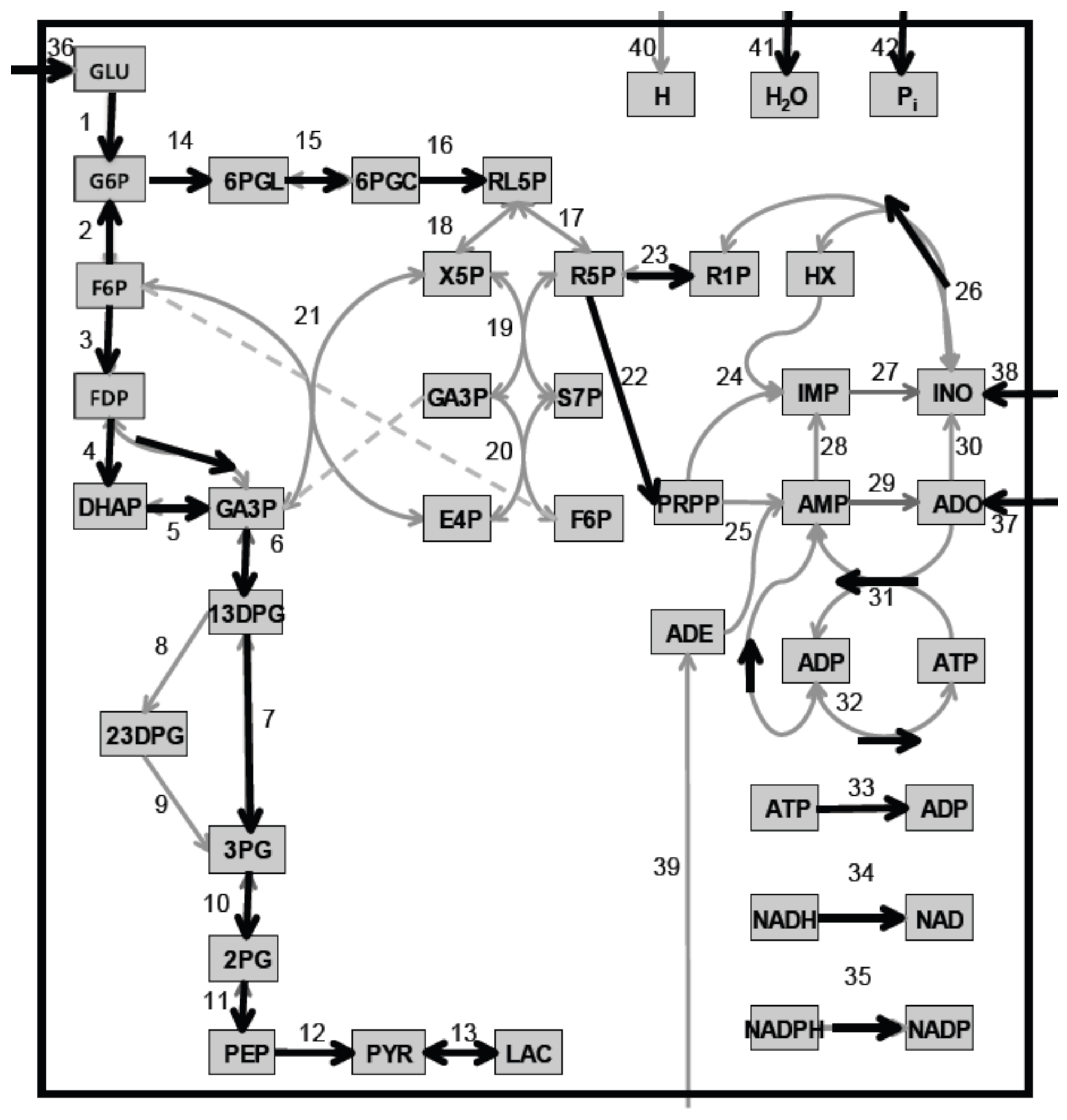}}
\caption{\label{corre}Large overlap backbone of hRBC: black arrows mark reactions with an overlap larger than 0.9 in an optimal healthy cell.}
\end{figure}

The flux distribution corresponding to optimal states in the healthy and enzyme deficient hRBC are displayed reaction by reaction in Fig.\ref{dis}, obtained by sampling 10000 solutions of (\ref{vn}), while a pictorial representation of the optimal flux states is given in Fig.\ref{map}.
For the healthy cell (black line in Fig.\ref{dis} and top left panel in Fig.\ref{map}) the large flux backbone is formed by the second part of the glycolysis (crucial for ATP, NADH and 23DPG production) and the PP pathway (NADPH production). The latter gives a substantial contribution to the former, not just as salvage way. The adenosine metabolism shows instead lower flux values. In addition to GLU, which is the fundamental substrate for hRBCs, the INO uptake plays an important role as an alternative way to the PP pathway. It is worth stressing that these solutions imply a net production of 23DPG, the crucial regulator for oxygen release, which is obtained without any imposed constraint. This picture is strongly reminiscent of the first eigenpathway obtained by extreme pathways analysis in \cite{pricea}, though the thermodynamic constraints and production requirements used in \cite{pricea}, including one on 23DPG, are more strict than the self-consistent analysis presented here. In Fig. \ref{corre} we report the overlap map of the hRBC. 
Comparing this with Fig. \ref{lupi} one sees that the large overlap backbone coincides to a large degree with the structurally most vulnerable parts of the network. Note that the overlap of reactions 2, 4, 15 and 26 is larger in the direction that cannot be replaced, further pointing to a higher susceptibility, and that currency reactions (31--35) belong to the most constrained part of the network. Revealingly, however, topological and dynamical characterizations prove to be complementary in some cases. This is seen e.g. from reaction 3, which is flux-constrained but also highly replaceable, so that the knockout damage is limited even in presence of a small allowed dynamical range. (A similar picture holds for reaction 23).
To conclude, we remind that in our framework uptake fluxes are optimized variables not fixed by boundary conditions. In the optimal state five of the uptakes have a limited allowed variability, implying rather severe constraints on the cell's environment.

\subsection{Response to enzymopathies}

We have simulated the most studied enzymopathies by constraining the flux of the corresponding reaction. Generically speaking, the hRBC metabolism displays a large resilience against partial perturbations. Indeed, we have observed appreciable differences in relevant cellular functions compared to the non-deficient case only under full enzyme deficiences, as also observed in \cite{sysbio} within a standard FBA optimization approach. Even under the most serious enzyme deficiencies the network appears to be able to maintain the production of ATP, NADH and NADPH almost constant, see also \cite{tekir}. We focus here on PK and G6PDH deletions. As shown in Fig.\ref{dis}, the alterations in the flux distributions are not particularly striking and indeed we do not observe global flux rearrangements on the network's scale. The G6PDH enzymopathy  appears to only cause local changes, confirming the structural predictions, the overlap calculations and also in agreement with clinical observations \cite{Jac}. The response to PK knockout is instead more marked. The synoptic analysis of Fig.\ref{map} shows that in general the response to the perturbation consisted in a drop of the GLU uptake, and in a reduction of the glycolitic flux, while the Rapoport-Leubering shunt (reactions 8-9) for the production of 23DPG remains particularly stable, as does the adenosine metabolism. For the glycolytic deficiences PK and HK we further observe an increase of the INO uptake to sustain the PP pathway and allow for the second part of glycolysis, and with it the production of ATP and NADH, to take place. Configurations corresponding to the next most severe enzymopathies (HK, EN, PGK and PGM) are available from http://chimera.roma1.infn.it.

\vspace{-0.5cm}

\section{Final remarks}

In this work we compare two robustness measures for biochemical networks, one based on structural properties (the reaction replaceability), the other based on dynamical capabilities (the overlaps). The latter depends on both the network topology and the model defined on it. Within VN's frame, we found that unreplaceable reactions largely correspond to processes with a smaller allowed flux variability. In such directions, reaction knockouts as well as constraints on the fluxes are mostly harmful. Reactions with limited (but non zero) replaceability tend to have instead smaller overlap, so that while the reaction is difficult to substitute still its flux can be largely adjusted. In an evolutionary perspective \cite{vit}, the former pathways appear as `frozen', and perturbations at these nodes will require large-scale flux rearrangement, while a mutation affecting the latter group may be neutral and could be preserved across generations. Interestingly, some reactions have both a large overlap and a large replaceability. These, albeit structurally robust, are dynamically constrained and should be considered as critical pathways of the metabolism as well. Integrating dynamical and structural characterizations thus provides a rather complete picture of the emerging network robustness. The structural analysis performed here was made possible by the small size of hRBC's metabolic network, which has served here as a model system to test basic concepts and algorithms. The use of the same procedure on a larger network, such as {\it E.coli}'s, is likely to be prevented by CPU time growth. However, message-passing algorithms, designed specifically to solve combinatorial optimization or counting problems -albeit approximately- on graphical models, may be a suitable replacement \cite{VKMa}. 

\vspace{-0.5cm}

\section*{Acknowledgment}
It is a pleasure to thank S. Jain for pointing us to \cite{samal} and D. Segr\`e for important comments and suggestions.

\vspace{-0.5cm}

%\paragraph{Funding\textcolon} Text Text Text Text Text Text  Text Text.

%\bibliographystyle{natbib}
%\bibliographystyle{achemnat}
%\bibliographystyle{plainnat}
%\bibliographystyle{abbrv}
\bibliographystyle{bioinformatics}
%
%\bibliographystyle{plain}
%
%\bibliography{Document}

\end{document}